\documentclass[twocolumn,showpacs,preprintnumbers,amsmath,amssymb,superscriptaddress,prb]{revtex4-1}
\usepackage{epsfig,amsopn}
\usepackage{graphicx}
\usepackage{epstopdf}
\usepackage{sidecap}
\usepackage{color}
\usepackage{subcaption}  
\usepackage{amsmath,amssymb}
\usepackage{amsthm}
\usepackage{enumerate}
\newcommand\beq{\begin{equation}}
\newcommand\eeq{\end{equation}}

\begin{document}
\title{Interplay of Floquet Lifshitz transitions and topological transitions in bilayer Dirac materials}
\author{Priyanka Mohan}
\affiliation{Department of Theoretical Physics, Tata Institute of Fundamental Research,
Homi Bhabha Road, Mumbai 400005, India }
\affiliation{Harish-Chandra Research Institute, HBNI, Chhatnag Road, Jhunsi, Allahabad 211 019, India.}
\author{Sumathi Rao}
\affiliation{Harish-Chandra Research Institute, HBNI, Chhatnag  Road, Jhunsi, Allahabad 211 019, India.}
\begin{abstract}
 
 We show how transitions between different Lifshitz phases in bilayer Dirac materials with and without spin-orbit coupling can be studied by driving  the system.  The periodic driving is induced by a laser and the resultant phase diagram is studied 
in the high frequency limit using  the Brillouin-Wigner perturbation  approach to leading order. The examples of such materials include bilayer graphene and spin-orbit coupled materials such as bilayer silicene.
The phase diagrams of the effective static models are analyzed to understand the interplay of topological phase transitions, with changes in the Chern number and
topological Lifshitz transitions, with the ensuing changes in the Fermi surface. Both the topological transitions and the Lifshitz transitions are  tuned by the amplitude of the drive.
\end{abstract}

\maketitle
\section{Introduction}

Bilayer graphene\cite{novoselov2005,novoselov2006,mccann2006,guinea2006,latil2006,partoens2006,partoens2007,koshino2009}  has been intensely studied in the last few years, mainly due to the tunability of its band-gap
by an electric field.
The two layers can be oriented differently, but what is most commonly studied is the AB or Bernal stacking, 
which is when the top layer is shifted so that half its atoms lie directly over the center of the hexagon of the lower graphene sheet. The band structure no longer has Dirac cones and the charge carriers in bilayer graphene are expected to be massive Dirac fermions. On the other hand, for spin-orbit coupled materials like silicene (and other  materials like germanene and stanene)\cite{ezawa2011,drummond2012,liu12011,liu22011,ezawa2015}, which have a buckled structure, tunability is already present in a monolayer, and it is well-known that the band structure can be tuned between band insulators and topological insulators through a metallic phase by applying an electric field. However, bilayer silicene\cite{ezawa2012,fu2014,padilha2015,zhang2015,do2017} (and other spin-orbit coupled materials) have even richer physical properties due to the interplay of buckling and stacking, and are beginning to be studied in detail. 

 Bilayer silicene is a trivial topological insulator, unlike monolayer silicene,
due to the lack of topologically protected edge states, in the presence of interlayer spin-orbit coupling and Rashba spin-orbit coupling.
 However, it has been argued that many of its physical properties
are similar to those of topological insulator and it should be considered a quasi-topological insulator\cite{ezawa2012}.
 At a  critical electric field, the quasi-topological phase of bilayer silicene makes a transition to a band insulator, but due
 to trigonal warping, there are several such critical electric fields  where the band closes, and the band structure is
 thus controllable  by the electric field to an even larger extent than in bilayer graphene. The electronic structure
 and properties of bilayer silicene are also strongly stacking dependent\cite{fu2014,padilha2015}  and recent work has
 also investigated the effect of perpendicular electric and magnetic fields on bilayer silicene.


The study of light-matter interaction has become increasingly popular in the last few years and  new developments in laser studies have led to the possibility of Floquet 
engineering\cite{goldman2014,bukov2015} which is the generation of new Hamiltonians that do not exist
in static systems.   In particular,  besides experiments in cold atom\cite{rechtsman2013} and photonic systems\cite{jotzu2014},
with the emergence of new topological phases in condensed matter systems\cite{wang2013},
the field of Floquet topological insulators\cite{oka2009,kitagawa2011,lindner2011,dora2012,rudner2013,kundu2014,usaj2014,titum2015,farrell2015,kundu2016,titum2016,mikami2016,mohan2016,klinovaja2016}  has shot into prominence in recent times.

 The possibility of tuning the band-gap in graphene and bilayer graphene using light greatly increases potential applications. It has also been shown\cite{varlet2015,shtyk2016} that bilayer graphene is an ideal system to study  Lifshitz transitions, because the parabolic dispersion in Bernal stacked bilayer graphene is not  protected by the crystal symmetry and is trigonally deformed at the lowest energies due to next nearest neighbour interlayer hopping, giving rise to four Dirac cones. Under an inter-layer voltage bias,
the Dirac points can move and cause a Lifshitz transition  - i.e., 
can change the topology of the Fermi surface.
In this paper, we study the effect of shining light on bilayer graphene as well as bilayer silicene and other spin-orbit coupled Dirac materials, in the high frequency limit using the Brillouin-Wigner perturbation approach. It was shown recently\cite{shtyk2016} that three van Hove saddles merge at a multi-critical Lifshitz point in bilayer graphene and it was shown that four different phases with different Fermi surface topologies occur.  One of the main results that we obtain in this paper is the effect of driving on the phase diagram with these four different Fermi surface topologies. The driving also induces new topological phases with different Chern numbers and we obtain those phases as well.
Although the effect of irradiation on bilayer graphene\cite{morell2012,dallago2017} has been recently studied, here we focus specifically on the the Lifshitz transitions and how they can be affected by driving. Moreover, we include effects of buckling as well as spin-orbit coupling so that other materials such as silicene, germanene, etc can also be incorporated, where  far fewer studies exist.

The plan of our paper is as follows. In Sec. II, we briefly review the Lifshitz transition in bilayer graphene to set the notation. In Sec. III, we include the intrinsic spin-orbit coupling and the buckling,
which are the hallmarks of silicene and other spin-orbit coupled materials and emphasize how the Lifshitz transition in these materials differs from graphene. In Sec. IV, we include periodic driving and
study the Brillouin-Wigner (B-W) perturbation expansion, without and with spin-orbit coupling, in bilayer systems, generalizing earlier results \cite{mikami2016,mohan2016} on single layers ands obtain the effective static Hamiltonian. 
In Sec V, we obtain the phase diagrams and show how the Lifshitz transition gets modified in the presence of light for both bilayer graphene and bilayer spin-orbit coupled materials.  We end with a small discussion in Sec. VI, where  we briefly discuss how the Lifshitz transitions may be observed in a real system.

\section{Bilayer graphene and the Lifshitz transition}
%
\begin{figure}[!ht]
 \centering
  \includegraphics[width=8cm]{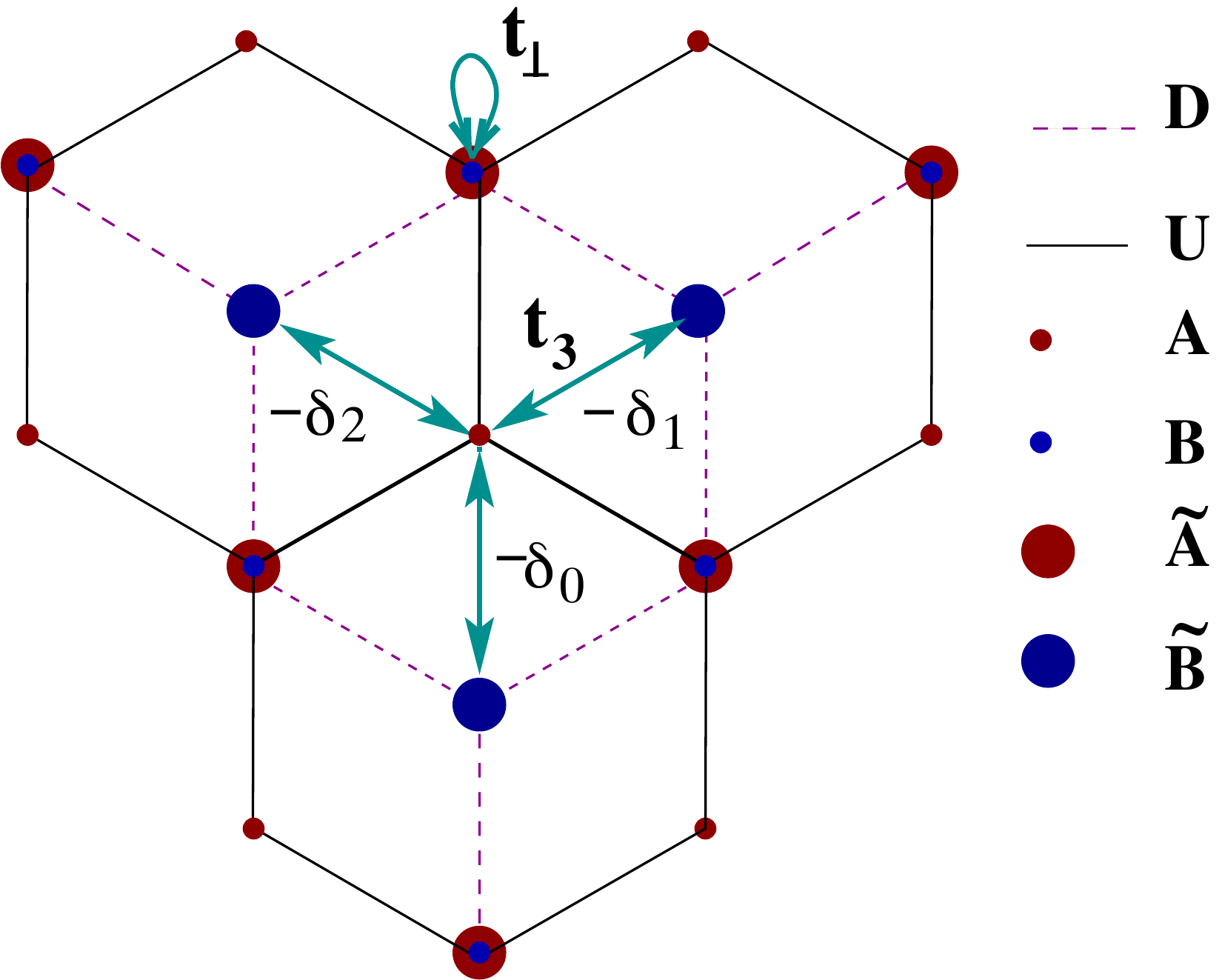}
  \caption{Bilayer model viewed from above. The small dots (red and blue)  denote the carbon atoms (on A and B sub-lattices) in one layer (U or up) whereas the big dots denote the same in the other layer (D or down).}
  \label{Fig:bilayer}
\end{figure}

The Hamiltonian of bilayer graphene can be written as the sum of the Hamiltonians of the individual layers and the inter-layer couplings between them. 
There are several different possible stackings including layers which are twisted with respect to one another. However, here we shall work with the most common Bernal or AB stacking. 
Although some details of the phase diagram may
change in the other cases, we expect that the general features of the phase diagram will essentially remain the same. The Hamiltonian for
Bernal stacking is given by
 \begin{equation}
  H_{BL}=H^U_{SL}+H^D_{SL}+H_{\text{inter}}
  \label{Eqn:Si_biH}
 \end{equation}
where the $H^U_{SL}$, $H^D_{SL}$ are the two single layer Hamiltonians for the up and down layers given by 
\begin{eqnarray}
 H_{\text{SLG}}=-t \sum_{\langle i j\rangle,\sigma} a^\dagger_{i,\sigma} b_{j,\sigma} + h.c
 \label{Eqn:SGH}
\end{eqnarray}
 and  $H_{\text{inter}}$ is the interlayer coupling. The index $\sigma$ refers to real spin and is usually dropped in the graphene context, since the Hamiltonian can be written independently for each spin. The sub-lattices of the two layers are denoted as 
 $A,B$ for the up layer and
$\tilde{A}, \tilde{B}$ for the down layer and the 
the Hamiltonian for the inter-layer hopping $H_{\text{inter}}$ is given by,
\begin{eqnarray}
 H_{\text{inter}}&=&t_\perp \sum_{i\in \tilde{A}, j\in B} \left( \tilde{a}^\dagger_{i\sigma}b_{j\sigma}
 +b^\dagger_{j\sigma}\tilde{a}_{i\sigma}\right)\nonumber\\
 &+& t_3 \sum_{i\in A, j\in \tilde{B}} \left(a^\dagger_{i\sigma} \tilde{b}_{j\sigma}
 + \tilde{b}^\dagger_{j\sigma} a_{i\sigma}\right)~.
 \label{Eqn:Hinter}
\end{eqnarray}
As shown in Fig.\ref{Fig:bilayer}, there are two distinct types of hopping terms between the layers. The hopping term $t_\perp$ is between sub-lattices $B$ and $\tilde{A}$  which are separated by $2L$ in the direction 
perpendicular to the planes, whereas the interlayer hopping term $t_3$ is between sub-lattices $A$ and $\tilde{B}$ separated by $2L$ in the perpendicular direction and also  one bond length along the planes. 
The magnitude of both interlayer hopping terms are $t_3, t_\perp \sim 0.1 t$, where $t_\perp\gtrsim t_3$.
Here, we have not considered the interlayer spin-orbit  coupling.

\begin{figure}[!ht]
   \includegraphics[width=9cm]{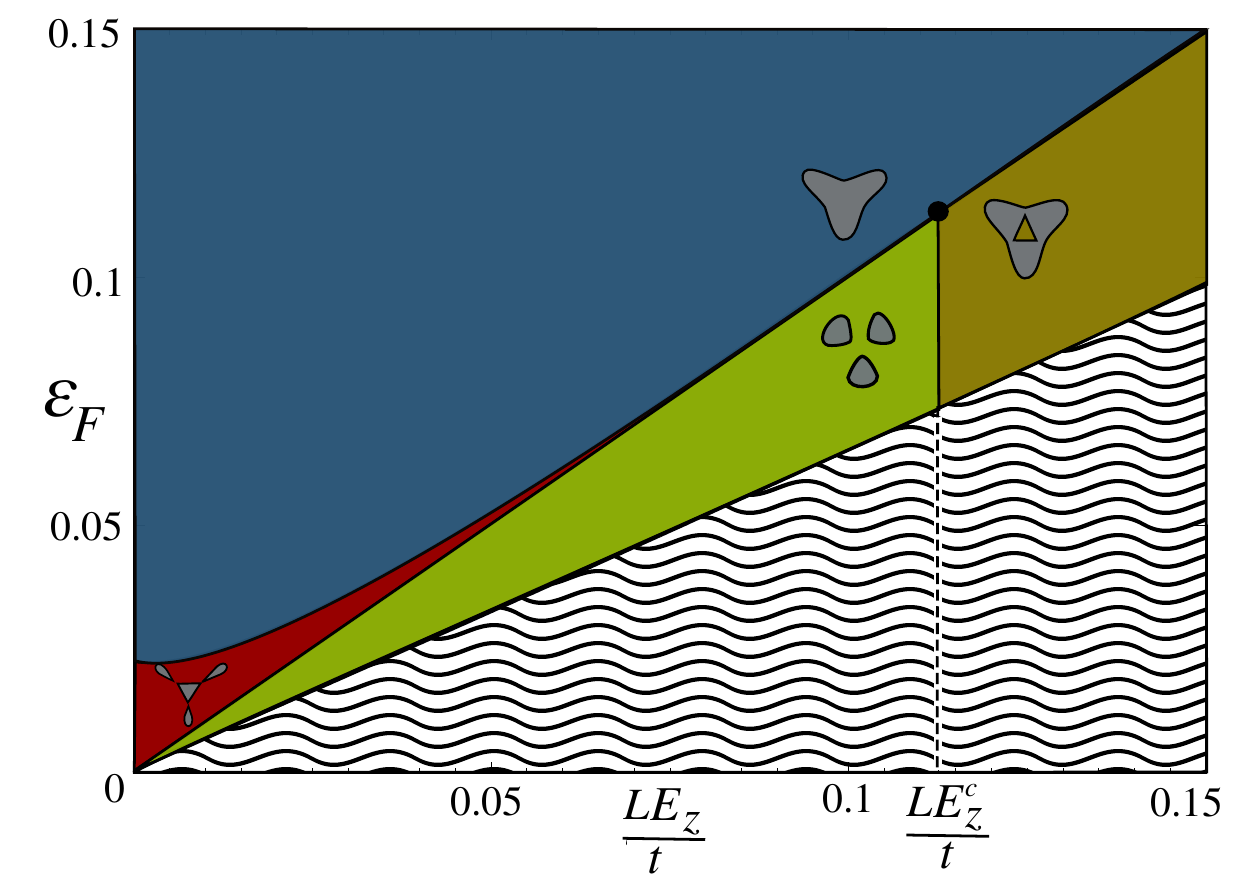}
   \caption{The Lifshitz phase diagram of bilayer graphene. The different phases are plotted with an increasing electric field. The shape of Fermi surface in each phase is shown in grey color.
   The black dot at $E^c_z$ is the Lifshitz multi critical point. The values of inter-layer couplings are $t_\perp=0.5t$ and $t_3=0.45t$ in this phase diagram for illustrative purposes.}
    \label{Fig:bare_lifshitz}
  \end{figure}

The total low-energy Hamiltonian in the vicinity of the Dirac points (for each spin) can be obtained in the momentum space and  is given by 
\begin{widetext}
 \begin{align}
 \psi^\dagger_{q} H_\eta \psi_q =\psi^\dagger_{q}\begin{pmatrix}
  LE_z &\frac{3a t}{2} \left(\eta q_x-iq_y\right)&0&-\frac{3a t_3}{2} \left(\eta q_x+iq_y\right)  \\
\frac{3a t}{2} \left(\eta q_x+iq_y\right) &  LE_z&t_\perp&0  \\
 0&t_\perp & - LE_z &\frac{3a t}{2} \left(\eta q_x-iq_y\right) \\
 -\frac{3a t_3}{2} \left(\eta q_x-iq_y\right) &0&\frac{3a t}{2} \left(\eta q_x+iq_y\right)  & - LE_z \\
 \end{pmatrix} \psi_q 
  \label{Eqn:H_matrix}
 \end{align}
\end{widetext}
where $\psi_{q}=\begin{pmatrix}
  a_{q}&b_{q} &\tilde{a}_{q}& \tilde{b}_{q}  \\
 \end{pmatrix}^T$, $\eta=\pm1$ for $K,K'$ is the pseudo-spin index and $a$ is the nearest-neighbour distance.
 An electric field added perpendicular to the layers will add a potential difference between the two layers.
The four eigenvalues of the above Hamiltonian are given by 
\begin{widetext}
\begin{eqnarray}
 \epsilon_{q,\eta}&=&\pm \frac{1}{\sqrt{2}}\left(2L^2E^2_z+\frac{9a^2q^2}{4}(2t^2+t^2_3)+t^2_\perp \right.\nonumber\\
 &\pm&\left.\left(t^4_\perp+9a^2 q^2 t^2(4L^2E^2_z+t^2_\perp)+\frac{9a^2q^2t^2_3}{4}(9a^2q^2(t^2+\frac{t^2_3}{4})-2t^2_\perp) 
  -27\eta a^3t^2t_3 t_\perp q_x(q^2_x-3q^2_y) \right)^{1/2}\right)^{1/2}
 \label{Eqn:bigev1}
\end{eqnarray}
\end{widetext}
where $q^2=q^2_x+q^2_y$. Precisely at the $K$ and $K^\prime$ points, the four eigenvalues reduce to 
\begin{eqnarray}
 &&\epsilon_{\eta}=\pm LE_z, ~ \pm \sqrt{L^2E^2_z+t^2_\perp}.
 \label{Eqn:epsiloneta}
\end{eqnarray}
 The linear dispersion of the two graphene layers become quadratic on the addition of the coupling $t_\perp$. Switching on the $t_3$ term adds trigonal warping to the bands near the $K(K')$ points. 
 This results in the quadratic band dividing into four Dirac cones: one at the $K(K')$ point and three satellite cones around it. The  three satellite Dirac points are situated at $\left(\frac{\eta t_\perp t_3}{3a t^2},\pm\frac{t_\perp t_3}{\sqrt{3}at^2}\right)$ and
  $\left(-\frac{2t_3t_\perp}{3a\eta t^2},0\right)$. These points can be easily computed by checking for nodes at non-zero momenta along the $q_y=0$ axis for $E_z=0$ and using the
  trigonal symmetry.  Each of the  satellite Dirac points is separated from the $K$ ($K^\prime$) point by a van-Hove singularity, which here is a maximum. At the satellite Dirac points, Eq.\ref{Eqn:bigev1} takes the form, 
  \begin{eqnarray}
   \epsilon_{\text{satellite}}&=&\pm\left(2L^2 E^2_z t^4+(t^2+t^2_3)^2 t^2_\perp \right. \nonumber\\
   &\pm& \left. t_\perp \left(16L^2 E^2_z t^6 t^2_3+(t^2+t^2_3)^4 t^2_\perp \right)^{1/2} \right)^{1/2}.
   \label{Eqn:epsilonsatellite}
  \end{eqnarray}
Note that two of the values in both Eq.\ref{Eqn:epsiloneta} and Eq.\ref{Eqn:epsilonsatellite} are zero  when $E_z=0$. As we increase the value of $E_z$, the system becomes gapped and the van-Hove singularity moves towards the 
$K$ ($K^\prime$) points. The four Dirac cones are replaced by electron-like pockets here. At a critical $E_z$ given by $LE^\text{critical}_z=t_\perp t_3/ 2t$, the van-Hove singularities merge 
causing a Lifshitz multi-critical point\cite{shtyk2016}. For $E_z>E^\text{critical}_z$, the central electron-like pocket becomes a hole-like one while the other three remain electron-like.  
In Fig.\ref{Fig:bare_lifshitz}, we show the phase diagram with phases with different Fermi surface topologies, similar to the one shown in Ref.\onlinecite{shtyk2016}. However, we choose
to show it as a function of the external perpendicular electric field $E_z$ and the Fermi energy. 
   
  To see the different Fermi surface topologies with an increasing electric field we have to include the chemical potential $\mu\,(\varepsilon_F)$ term  in Eq.\ref{Eqn:bigev1} where it has been set to zero.
 The wavy striped region in Fig.\ref{Fig:bare_lifshitz} is when the  Fermi energy or chemical potential lies in the gap between the conduction and valence bands. An increase in the electric field pushes the valence and 
 conduction bands away from each other increasing the striped region. For $E_z=0$,  shifting the Fermi level upwards  will result in four disconnected Fermi surfaces, one from the Dirac cone and one from each of the three satellite 
 cones. This phase forms the red region in the phase diagram. As we increase the external electric field from zero, the distance between the valence and conduction bands at $K/K'$ points
 increases further than at the 
 satellite points.  When the Fermi level is between the Dirac cone and the satellite cones, the Fermi surface consists of three disconnected regions. This area of the phase diagram is shown in 
 green. The height of the line (measured from zero) separating  the blue and red regions in the phase diagram is the height of the van-Hove singularity measured from the point half-way between the valence and conduction bands. 
 Above this phase boundary, the Fermi energy has increased to the level that 
  the Dirac cone and the satellite cones merge to  a single band with the topology of a single region. As we increase the external electric field, the  three van-Hove singularities move closer to 
 the $K/K'$ point and merge at a critical electric field $E^c_z$. A multi-critical Lifshitz point occurs in the phase diagram at the $E^c_z$ where all the different Lifshitz phases meet. A further increase
 in electric field causes the electron-like pocket at the $K/K'$ point to become hole-like. This phase is the mustard region with the topology of an annulus. The Lifshitz phase transition between the blue and 
 the red phases in the phase diagram is of the 'neck-narrowing type'. As the name indicates, here the Fermi surface get a new topology by pinching off a region. The transition between the green 
 and mustard region is also of the same type. The phase transitions between the red and green phases and the one between the blue and mustard phases is of the 'pocket vanishing type'. Here the 
 topology of the Fermi surface changes when an electron-like pocket becomes hole-like or vice versa.



 \section{Extension to spin-orbit coupled bilayer materials}

 The difference between graphene and other two-dimensional (2D) materials like silicene or germanene is essentially due to the larger size of their atoms  which causes buckling of 
 the lattice as well as leading to the presence of an intrinsic spin-orbit coupling. In this paper, we study 
 a general material with arbitrary values for spin-orbit coupling and  buckling distance, so that the results can be applied to any of these materials. In the rest of the work, the term silicene 
 is used to represent such a general material. 
 
 \begin{figure}
  \centering
  \vspace{0.3cm}
 \includegraphics[width=7cm]{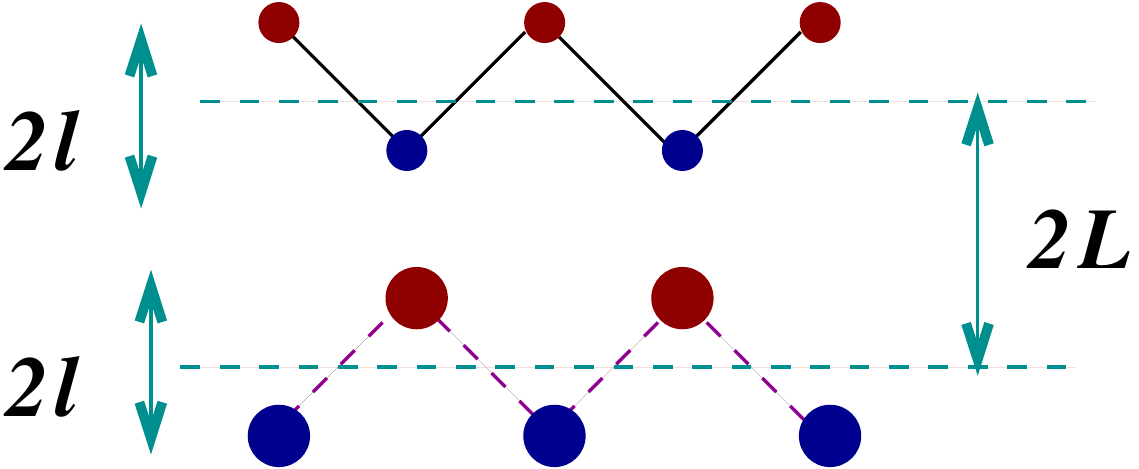}
 \caption{The different sub-lattices in bilayer silicene viewed horizontally layer by layer. The small and big dots denote the carbon toms in the upper (U) and lower (D) layers respectively.}
 \label{Fig:buck}
\end{figure}
 
 For a single layer, the  Hamiltonian of these materials  has additional terms beyond the terms in Eq.\ref{Eqn:SGH} given by, 
 \begin{eqnarray}
  H_{\text{SLS-extra}} &=&
\frac{i\lambda\sigma}{3\sqrt{3}}\sum_{\langle\langle i j \rangle \rangle,\sigma} \nu_{ij} (a^\dagger_{i,\sigma} a_{j,\sigma}+b^\dagger_{i,\sigma} b_{j,\sigma})\nonumber\\
  &+& \sum_{i,\sigma} lE_z  (a^\dagger_{i,\sigma} a_{i,\sigma}-b^\dagger_{i,\sigma} b_{i,\sigma}).
  \label{Eqn:SSH}
 \end{eqnarray}
 Here, $\lambda$ is the spin-orbit coupling between the sites of the same sub-lattice and $2l$ is the buckling distance as shown in Fig.\ref{Fig:buck}. The spin index $\sigma=\pm1$ 
corresponding to $ \uparrow/\downarrow$  and $\nu_{ij}=\pm1$ depending on whether the path taken from $j$ to $i$ is clockwise/counter-clockwise. An external electric field $E_z$ is added perpendicular to
the layers which creates  a potential difference between the $A$ and $B$ sub-lattices and also between the layers.

 The Hamiltonian of bilayer silicene has  earlier been studied in Ref.\onlinecite{ezawa2012}
 and its band structure and edge modes were obtained.   
 Here, we review this for two reasons - first, to understand Lifshitz transitions in this model which
 have not been studied earlier and second, to set our notation for the coupling to photons in the next section.
 We have two copies of the single layer Hamiltonian given by
 \beq
 H_{SLS}  =  H_{SLG} + H_{SLS-extra}
 \eeq
   and the inter-layer  coupling as given in Eq.\ref{Eqn:Hinter}. Like the case of
 bilayer graphene,  there are  two hopping terms between the layers which are separated by a distance $2L$ (see Fig.\ref{Fig:buck}). The hopping term $t_\perp$ is between sub-lattices $B$ and $\tilde{A}$  which are separated by $2(L-l)$ in the direction 
perpendicular to the planes. The other interlayer hopping term $t_3$ is between sub-lattices $A$ and $\tilde{B}$ separated by $2(L+l)$ in the perpendicular direction. Due to the individual 
buckling of both layers, all the four sub-lattices in bilayer silicene are on four different planes and therefore at four different potentials when $E_z$ is applied.
The magnitude of both interlayer hopping terms $t_3,t_\perp$ are $\sim 0.1 t$, and
furthermore, $t_\perp\gtrsim t_3$.  Note that the  interlayer spin-orbit  coupling has been set to zero, so
the model still has spin conservation. However, this conservation is not protected by a symmetry and so bilayer silicene is a `quasi-topological insulator'\cite{ezawa2012}.

 \begin{figure}[!ht]
  \includegraphics[width=9cm]{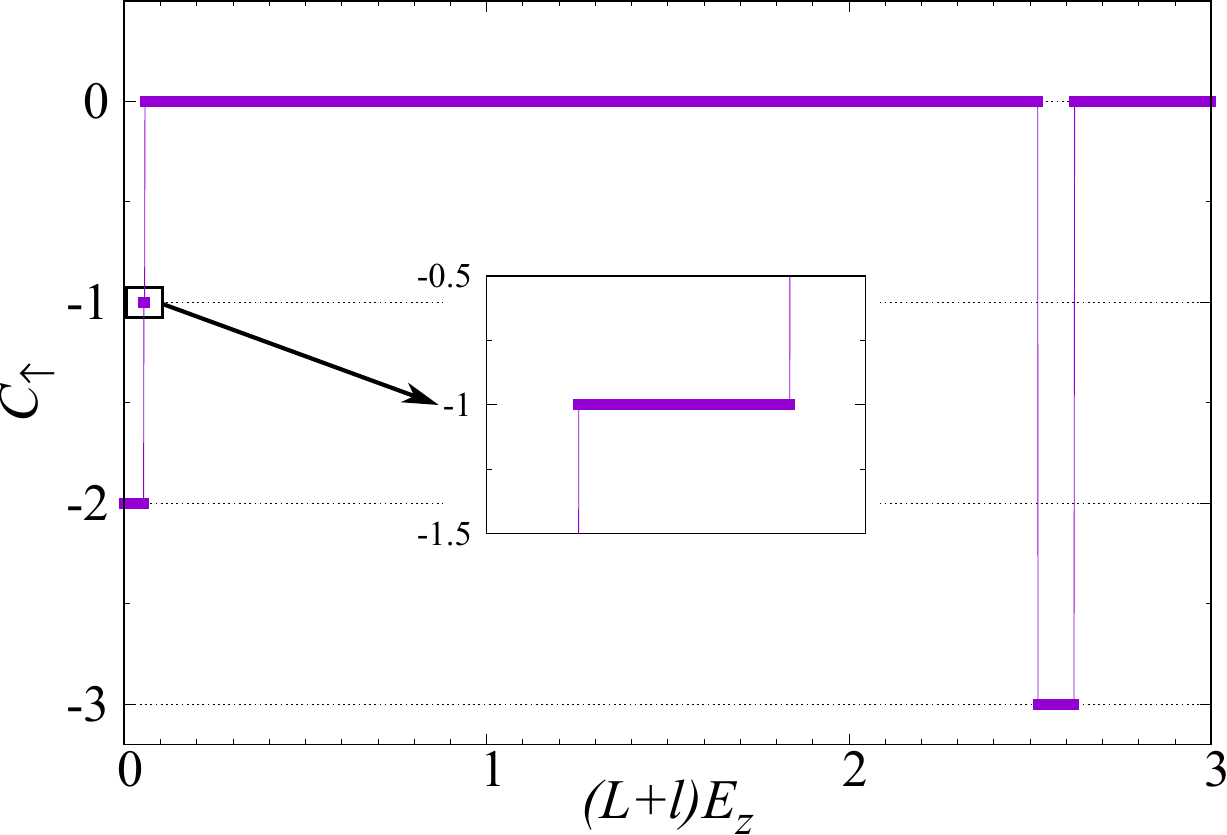}
  \caption{The Chern number, $C_\uparrow$ plotted as a function of  $(L+l)E_z$ for $\uparrow$  spin $(\sigma=1)$ in bilayer silicene. Due to the time-reversal symmetry of the model, $C_\downarrow=- C_\uparrow$. 
  The values of inter-layer couplings are $t_\perp=0.12t$, $t_3=0.1t$ and $l=L/4$.}
  \label{Fig:bilayersi}
\end{figure}

\begin{figure*}
 \centering
 \hspace{-1cm}
 \includegraphics[height=5cm,width=18cm]{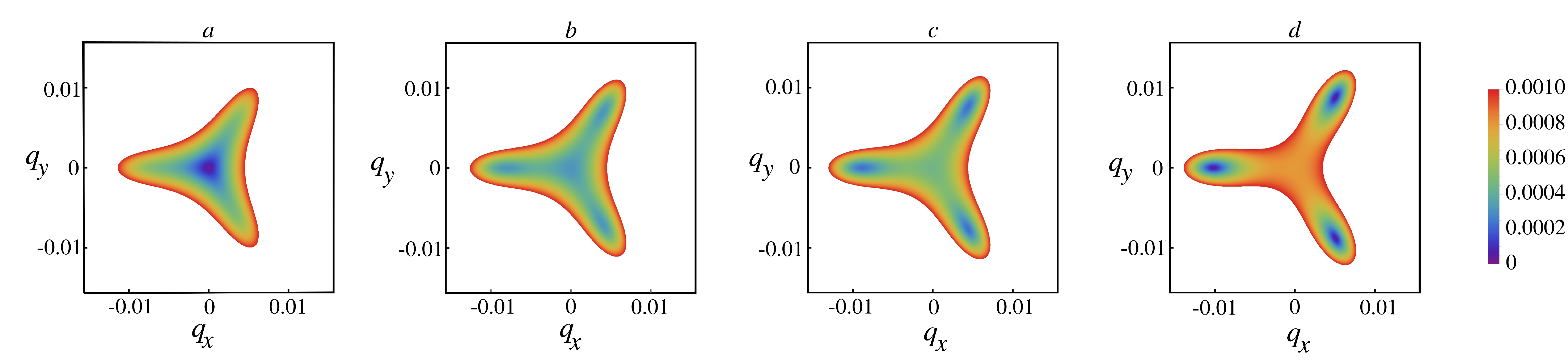}
 \caption{The plots from left to right corresponds to panels (a)-(d) in Fig.\ref{figbisi} and shows the evolution
 from a single Dirac cone at $K$ point to three Dirac cones on tuning the external electric field. This region is marked by the Chern number $-1$ in the inset of Fig.\ref{Fig:bilayersi} }
\label{bisiFST}
\end{figure*}

 The low-energy Hamiltonian expanded around the $K$ and $K^\prime$ points in the Brillouin  zone then reads,
\begin{widetext}
 \begin{align}
  H_\eta =\psi^\dagger_{q}\begin{pmatrix}
  \eta\sigma\lambda+(L+l)E_z &\frac{3a t}{2} \left(\eta q_x-iq_y\right)&0&-\frac{3a t_3}{2} \left(\eta q_x+iq_y\right)  \\
\frac{3a t}{2} \left(\eta q_x+iq_y\right) &   -\eta\sigma\lambda+ (L-l)E_z&t_\perp&0  \\
 0&t_\perp &  \eta\sigma\lambda -(L-l)E_z &\frac{3a t}{2} \left(\eta q_x-iq_y\right) \\
 -\frac{3a t_3}{2} \left(\eta q_x-iq_y\right) &0&\frac{3a t}{2} \left(\eta q_x+iq_y\right)  &  -\eta\sigma\lambda -(L+l)E_z \\
 \end{pmatrix} \psi_q 
  \label{Eqn:Hbisi_matrix}
 \end{align}
\end{widetext}
with the same $\psi_q,\psi_q^\dagger$ defined earlier. Note that because of the spin-orbit coupling, 
this Hamiltonian has
an explicit dependence on the spin index $\sigma$. So for each value of  $\sigma=\uparrow, \downarrow$, we get a specific $H_\eta$.  
Note  also that this reduces to the Hamiltonian in Ref.\onlinecite{ezawa2012} in the appropriate limit.

We can now obtain the eigenvalues of this Hamiltonian as 
\begin{widetext}
\begin{eqnarray}
 \epsilon_{q,\eta,\sigma}&=&\pm \frac{1}{\sqrt{2}}\left(2(L^2+l^2)^2E^2_z+\frac{9a^2q^2}{4}(2t^2+t^2_3)+t^2_\perp+2\lambda^2 +4lE_z\eta \sigma \lambda 
\right.\nonumber\\
&\pm& \left.\left(t^4_\perp+9a^2 q^2 t^2(4L^2E^2_z+t^2_\perp)+\frac{9a^2q^2t^2_3}{4}(9a^2q^2(t^2+\frac{t^2_3}{4})-2t^2_\perp) 
  -27\eta a^3t^2t_3 t_\perp q_x(q^2_x-3q^2_y) \right.\right.\nonumber\\
  &+&\left.\left.4LE_z (\eta\sigma\lambda+lE_z) \left(\frac{9}{2}a^2q^2t^2_3-2t^2_\perp\right)+16L^2E^2_z(\eta\sigma\lambda+lE_z)^2\right)^{1/2}\right)^{1/2}
 \label{Eqn:bisiev1}
\end{eqnarray}
\end{widetext}

\begin{figure}[!ht]
   \includegraphics[width=9cm,height=6cm]{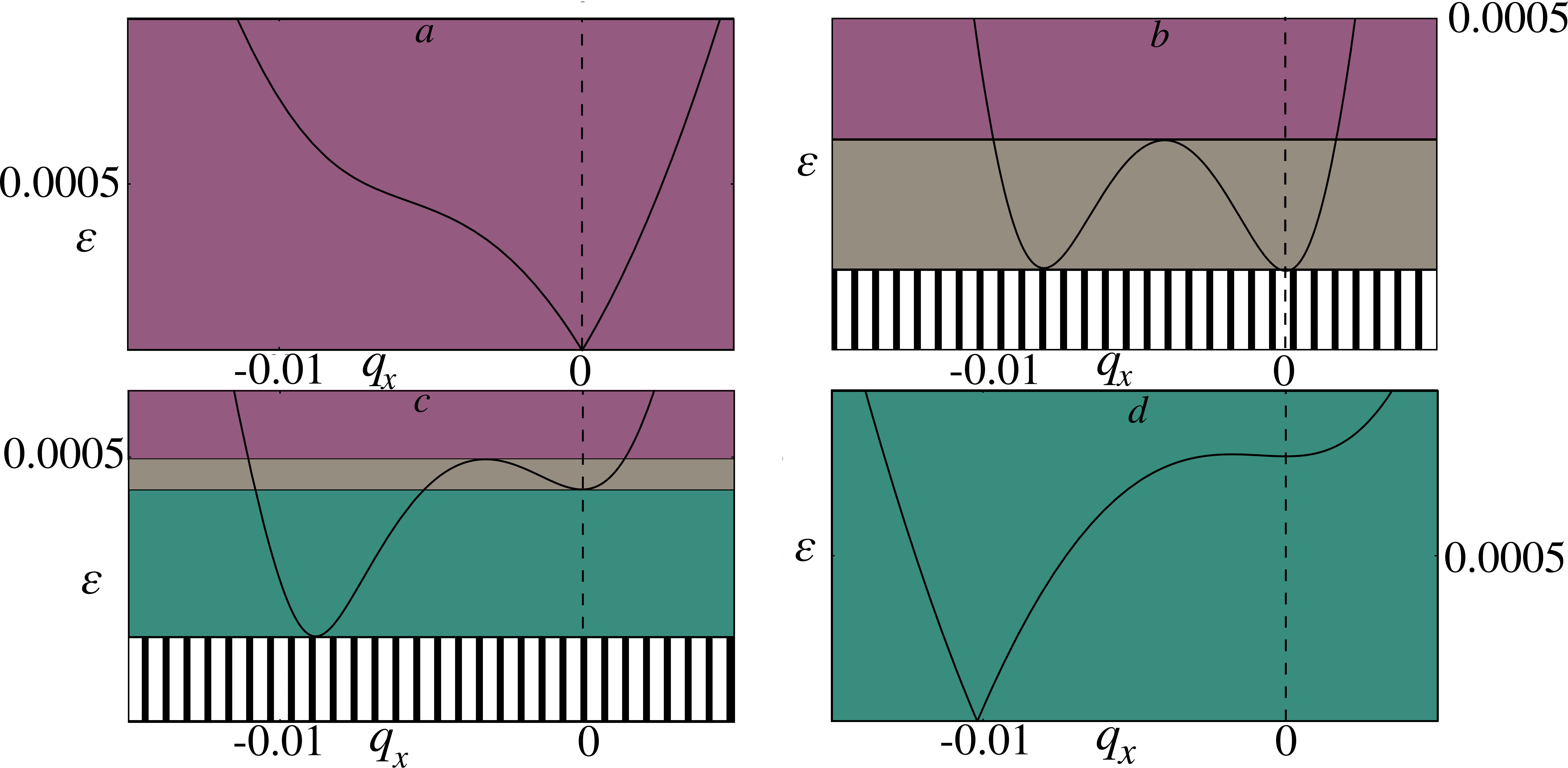}
   \caption{ The Lifshitz phase diagram of bilayer silicene. The four panels show the  evolution of the band structure on the $q_y$ axis (for $q_x=0$) near $K$ as the electric field $E_z$ is varied. 
   Panels (a) and (d)  have the band crossings at $K$ and satellite Dirac point respectively, while panels (b)-(c) are band structure evolutions between them. The different Lifshitz phases 
   for varying chemical potential values are marked by different colored regions in the phase diagram. }
%
    \label{figbisi}
  \end{figure}

Comparing Eqs. \ref{Eqn:bigev1} and \ref{Eqn:bisiev1}, we note that they differ by terms proportional to the buckling as well as by the spin-orbit coupling terms. Moreover, the eigenvalues are 
explicitly spin-dependent. At the $K/K'$ points, the eigenvalues have the form, 
\begin{eqnarray}
 \epsilon_{\eta,\sigma}&=&\pm (\eta\sigma\lambda+(L+l)E_z),\nonumber\\
 &&\pm \sqrt{(\eta\sigma\lambda-(L-l)E_z)^2+t^2_\perp}.
 \label{Eqn:bisiev2}
\end{eqnarray}
The most important difference between bilayer graphene and silicene is the gap due to the spin-orbit coupling. Like in the case of monolayer silicene, this gap 
can be closed by tuning the external 
electric field and leading to Chern number changes which are  depicted in Fig.\ref{Fig:bilayersi}.
As the strength of the external electric field increases from zero, the Chern number changes four times corresponding to four gap closings in the Brillouin zone. The first change in the 
 Chern number occurs when the gap closes at $K/K^\prime$ points when $(L+l)E_z=-\eta\sigma\lambda$, evident from Eq.\ref{Eqn:bisiev2}. The Chern number changes
  from $-2$ to $-1$ here. The second one occurs when the gap closes at the three (expected from trigonal symmetry) satellite Dirac points around
  $K/K^\prime$ points, as $E_z$ increased slightly. The Chern number goes from $-1$ to $0$ as shown in the inset of Fig.\ref{Fig:bilayersi}. The evolution of the  
  bands from when the gap closes  at the $K$ point (for the $\uparrow$ spin sector)  to when the gap closes  at the satellite Dirac points as a function of increasing $E_z$ is depicted in Fig.\ref{bisiFST}. This is also the region where 
  we choose to study the Lifshitz transition in this model.
  The third and fourth gap closings happens for much larger values of $E_z$ elsewhere in the Brillouin zone which are not accessible in the low energy Hamiltonian in Eq.\ref{Eqn:Hbisi_matrix}.

In Fig.\ref{figbisi}, we numerically show the evolution of the gaps at the $K$ and at the associated satellite point situated along the $q_y=0$ axis.
 The two other satellite Dirac points are found using the trigonal symmetry.
We now try to understand the Lifshitz transition  in bilayer silicene and the changes that occur in the Fermi surface topology. We have chosen to show the results for the spin $\uparrow$ sector. The results
for the spin $\downarrow$ sector are qualitatively similar. 

 When the chemical potential is set to zero along with the applied electric field the system has a gap of the $O(\lambda)$. On increasing the strength of the applied electric field, the gap closes 
at the Dirac point. This gapless band is plotted in Fig.\ref{figbisi}(a). By increasing the Fermi level from zero, we obtain a Fermi surface with a disc topology or topology of a single region. This Lifshitz phase is 
marked by purple colored regions in Fig.\ref{figbisi}(a)-(c). Increasing the electric field further gaps out the Dirac cone at $K$ point (for $\mu=0$) while simultaneously creating electron pockets at
the three satellite Dirac points. Fig.\ref{figbisi}(b)-(c) depicts 
two such cases. Depending on the value of Fermi energy we can get two additional Lifshitz phases along with an insulating phase here. The striped regions in Fig.\ref{figbisi}(b)-(c)
are insulating phases where Fermi energy lies below both the electron pockets. The brown regions are the phases with four disconnected Fermi surfaces and the green 
ones are the phases with three disconnected Fermi surfaces. Further increase of $E_z$ results in the gap closing at the satellite Dirac points as shown in Fig.\ref{figbisi}(d). A non-zero 
chemical potential here will give a Fermi surface with three disconnected regions.


\section{Periodic driving and the effective Hamiltonian for bilayer materials}
\label{Sec.BW}
In this section, we include the effects of periodic driving  in the high frequency limit, by obtaining an effective static Hamiltonian using the Brillouin-Wigner (B-W) perturbation theory. Generalizing earlier work on single layer systems like graphene\cite{mikami2016} and silicene\cite{mohan2016}, we find that 
for the bilayer Hamiltonian described in Eq. \ref{Eqn:Si_biH}, up to $O(1/\omega)$, the  Hamiltonian is  given by
\beq 
H^{\text{B-W}} = H^{(0)} + H^{(1)}
\eeq
where $H^{(0)}$ and $H^{(1)}$ are the zeroth and first order terms in the B-W expansion. The effect of radiation is taken into account by the vector
potential ${\bf A}({\tau}) = A_0(\cos\omega\tau, \sin\omega\tau)$, where $\omega$ is the frequency and we work in the high frequency limit. 
The hopping term is then modified by the Peierls substitution and changes to 
$-t \sum_{\langle i,j\rangle} e^{-i\alpha\sin(\omega\tau-2\pi l/3)} a_i^\dagger b_j$, 
where $l=0,1,2$ for the three nearest neighbours  in the honeycomb lattice.  We will generically include both buckling and the spin-orbit terms and
then set them to zero when considering bilayer graphene and use the appropriate values for the different spin-orbit coupled materials. The expansion has terms re normalizing the intra-layer couplings of  both the layers and the inter-layer terms between them. 
 The renormalization of the in-plane terms are
given by\cite{}
\begin{widetext}
 \begin{align} 
 \mathcal{J}_{\sigma} =& -t J_0(\alpha)+\frac{4t\sigma \lambda }{3\omega}\sum_{n\ne0}\beta_n\sin{\frac{\pi n}{6}} +\frac{t^3}{\omega^2} \left[\sum_{n\ne0} \gamma_n\left(2\cos{\frac{2\pi n}{3}+3}\right) +\sum_{m,n \ne 0}\chi_{nm}\left( 4\cos{\frac{2\pi n}{3}+1} \right)\right],  \nonumber   \\
 \Lambda^0_{\sigma} =& \frac{\sigma \lambda J_0(\alpha\sqrt{3})}{3\sqrt{3}}-\sum_{n\ne0} 
 \frac{t^2J^2_n(\alpha)}{\omega n}\sin{\frac{2\pi n}{3}}
 , \nonumber \\
 L_{\sigma} =&-\frac{4t\sigma\lambda}{3\omega}\sum_{n\ne0} \beta_n\sin{\frac{\pi n}{2}} +\frac{2t^3}{\omega^2} \left(\sum_{n\ne0} \gamma_n\cos{\frac{2\pi n}{3}} +\sum_{m,n \ne 0}\chi_{nm} \cos{\frac{2\pi(m-n)}{3}} \right), \label{eq:Lcoup} \nonumber \\
 M_{\sigma}=& -\frac{2t\sigma\lambda}{3\omega}\sum_{n\ne0} \beta_n\cos{\pi n}\,\sin{\frac{\pi n}{6}}
 +\frac{t^3}{\omega^2} \left(\sum_{n\ne0}\gamma_n\cos{\frac{2\pi n}{3}} 
 +  \sum_{m,n \ne 0}\chi_{nm} \cos{\frac{2\pi(m+n)}{3}} \right) ,   
  \end{align}
\end{widetext}
where the original parameters $t$ and  $\lambda$ are defined in Eqs. \ref{Eqn:SGH} and \ref{Eqn:SSH} and $\alpha=aA$. $J_n$ is the Bessel function of order $n$ and  $\beta_n = J_n(\alpha)J_n(\alpha\sqrt{3})/\sqrt{3}n$, $\gamma_n = J^2_n(\alpha)J_0(\alpha)/n^2$ and $\chi_{nm}=J_m(\alpha)J_n(\alpha)J_{m+n}(\alpha)/mn$. 

\begin{figure}[!ht]
 \centering
  \includegraphics[width=8cm]{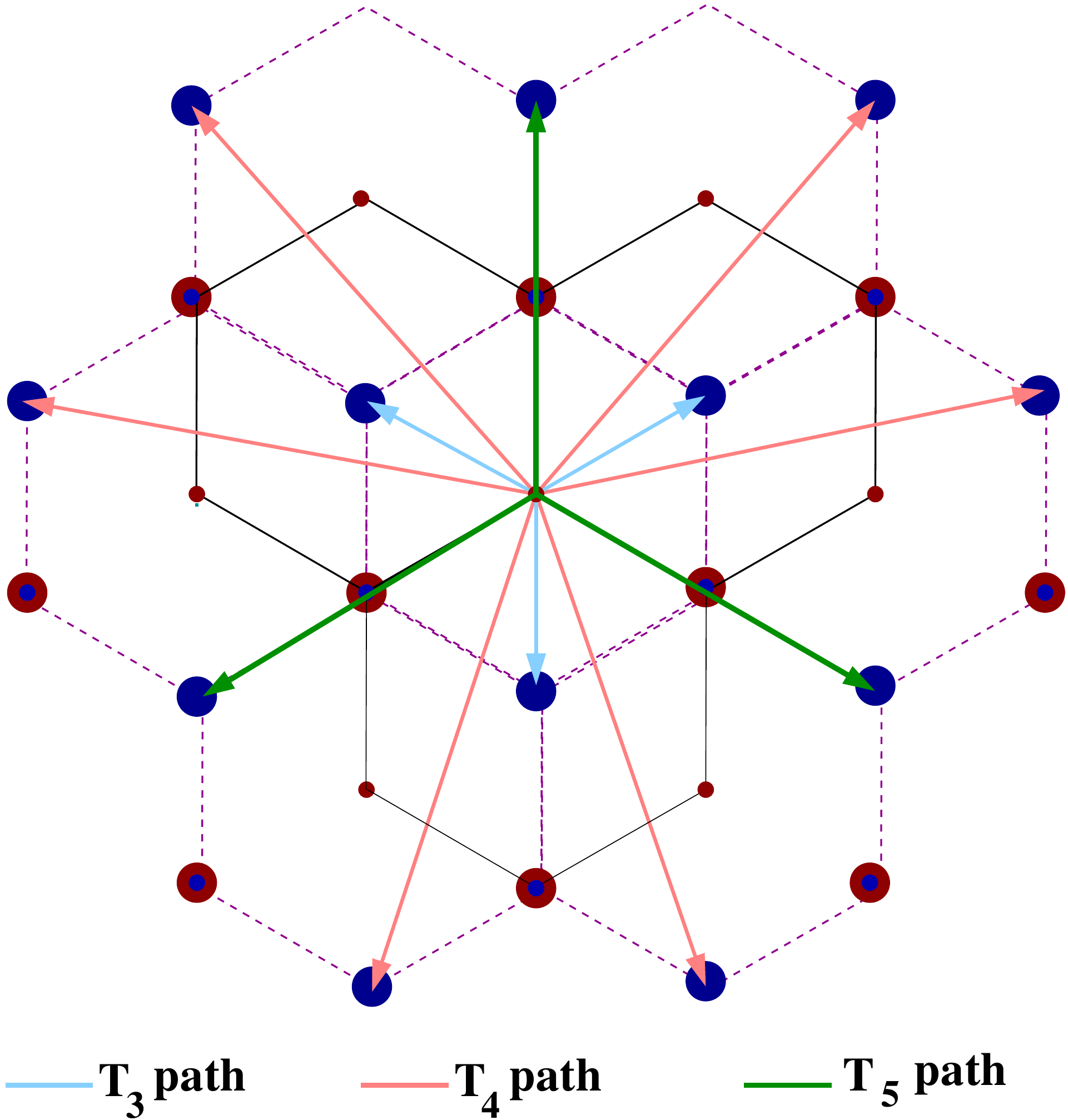}
  \caption{Different hopping paths between layers  in bilayer silicene}
  \label{Fig:bilayerh}
\end{figure}
The inter-layer Hamiltonian $H_\text{inter}$ can also be expanded order by order, leading to longer ranged 
hoppings as shown in Fig.\ref{Fig:bilayerh}.
The zeroth order term is given by
\begin{eqnarray}
 H^{(0)}_{}&=& t_\perp \sum_{i\in \tilde{A}, j\in B} \left( \tilde{a}^\dagger_{i\sigma}b_{j\sigma}
 +b^\dagger_{j\sigma}\tilde{a}_{i\sigma}\right) \nonumber\\
&+& t_3 J_0(\alpha)\left(\sum_{i\in A}  a^\dagger_{i\sigma} \tilde{b}_{i-\delta_l\,\sigma}
 + \sum_{j\in \tilde{B}}  \tilde{b}^\dagger_{j\sigma} a_{j+\delta_l\,\sigma}\right).\qquad
 \label{Eqn:bwzero}
\end{eqnarray}
 The $t_\perp$ term  exists only in the zeroth order term and does not occur at $O(1/\omega)$.

The first order term, $O(1/\omega)$ involving $t_3$ is given by,
\begin{eqnarray}
 H^{(1)}_{BW}&=&\sum_{\{n_i\}\ne0} \frac{H_{0,n_1}H_{n_1,0}}{n_1 \omega} \nonumber \\
 &=&\sum_{\{n\}\ne0} \frac{1}{n\omega}\left[T^{3}_{-n}T^{3}_{n} +T^{}_{-n}\, T^{3}_n+ \tilde{T}_{-n}\, T^{3}_n+
 T^{3}_{-n}\, \tilde{T}^{}_n  \right.\nonumber\\
&&\left.+T^{3}_{-n} \, T^{}_n \right].
\label{Eqn:bw1big}
\end{eqnarray}
Here, the terms of $O(tt_3)$ cancel  and 
the terms of $O(t^2_3)$  renormalizes  the spin-orbit coupling terms in  the $A$ sub-lattice of the top layer and the $\tilde{B}$ sublattice of the 
bottom layer as follows -
\begin{eqnarray}
 \Lambda^1=-i\frac{t^2_3 J^2_n(\alpha)}{n\omega}\sin{\frac{2n\pi}{3}} \sum_{\langle \langle i,j \rangle \rangle} \nu_{ij} (a^\dagger_i a_j +\tilde{b}^\dagger_i \tilde{b}_j ).
\end{eqnarray}

The terms $O(\lambda t_3)$  contribute the following three terms to the effective $H$ -
 \begin{eqnarray}
 T^\prime_3&=&\frac{4 t_3 \lambda \sigma}{3\sqrt{3}\omega n} J_n(\alpha\sqrt{3})J_n(\alpha)\sin{\frac{n\pi}{6}}
 \sum^{T_3-\text{path}}_{\substack{i\in A\, j \in \tilde{B}\\ \langle \langle \langle i,j \rangle \rangle \rangle}}  a^\dagger_i \, \tilde{b}_{j}\nonumber\\
T_4&=&\frac{-2 t_3 \lambda \sigma}{3\sqrt{3}\omega n} J_n(\alpha\sqrt{3})J_n(\alpha)\sin{\frac{n\pi}{6}} \cos{n\pi}
 \sum^{T_4-\text{path}}_{\substack{i\in A\, j \in \tilde{B}\\ \langle \langle \langle i,j \rangle \rangle \rangle}}  a^\dagger_i \, \tilde{b}_{j}\nonumber\\
T_5&=&\frac{-4 t_3 \lambda \sigma}{3\sqrt{3}\omega n} J_n(\alpha\sqrt{3})J_n(\alpha)\sin{\frac{n\pi}{2}}
 \sum^{T_5-\text{path}}_{\substack{i\in A\, j \in \tilde{B}\\ \langle \langle \langle i,j \rangle \rangle \rangle}} a^\dagger_i \, \tilde{b}_{j}
 \end{eqnarray}
The $T^\prime_3$ renormalizes the $t_3$ term in the bare Hamiltonian, whereas the $T_4$ and $T_5$ couplings are longer ranged inter-layer  hopping terms depicted in Fig.\ref{Fig:bilayerh}.

 \begin{figure}[!ht]
   \includegraphics[width=9cm,height=7cm]{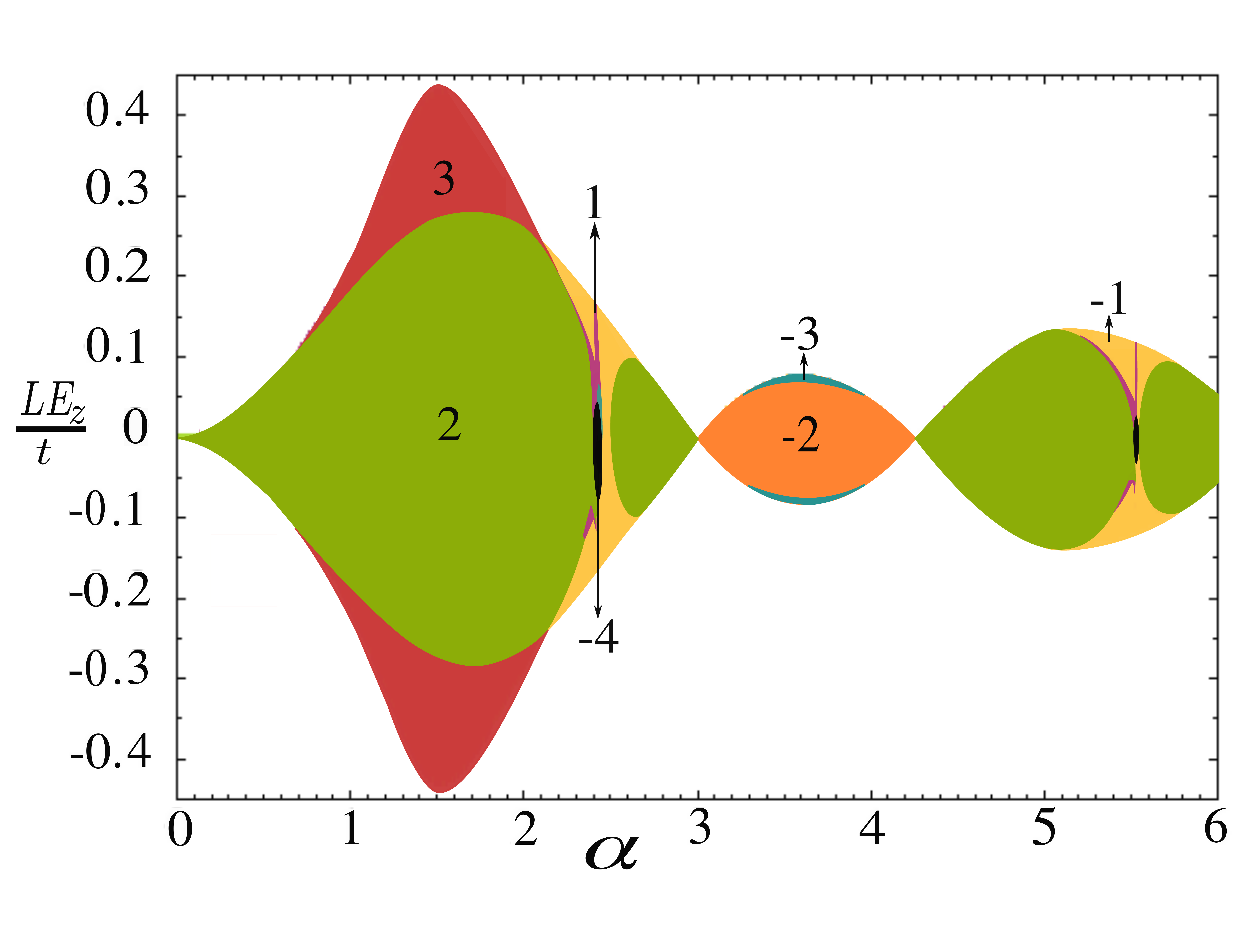}
   \caption{The Floquet topological phase diagram of bilayer graphene as a function
   of the drive amplitude $\alpha$. }
   \label{Fig:big_pd} 
  \end{figure}

\begin{figure*}
\centering
        \begin{subfigure}[b]{0.33\textwidth}
                \includegraphics[width=6cm]{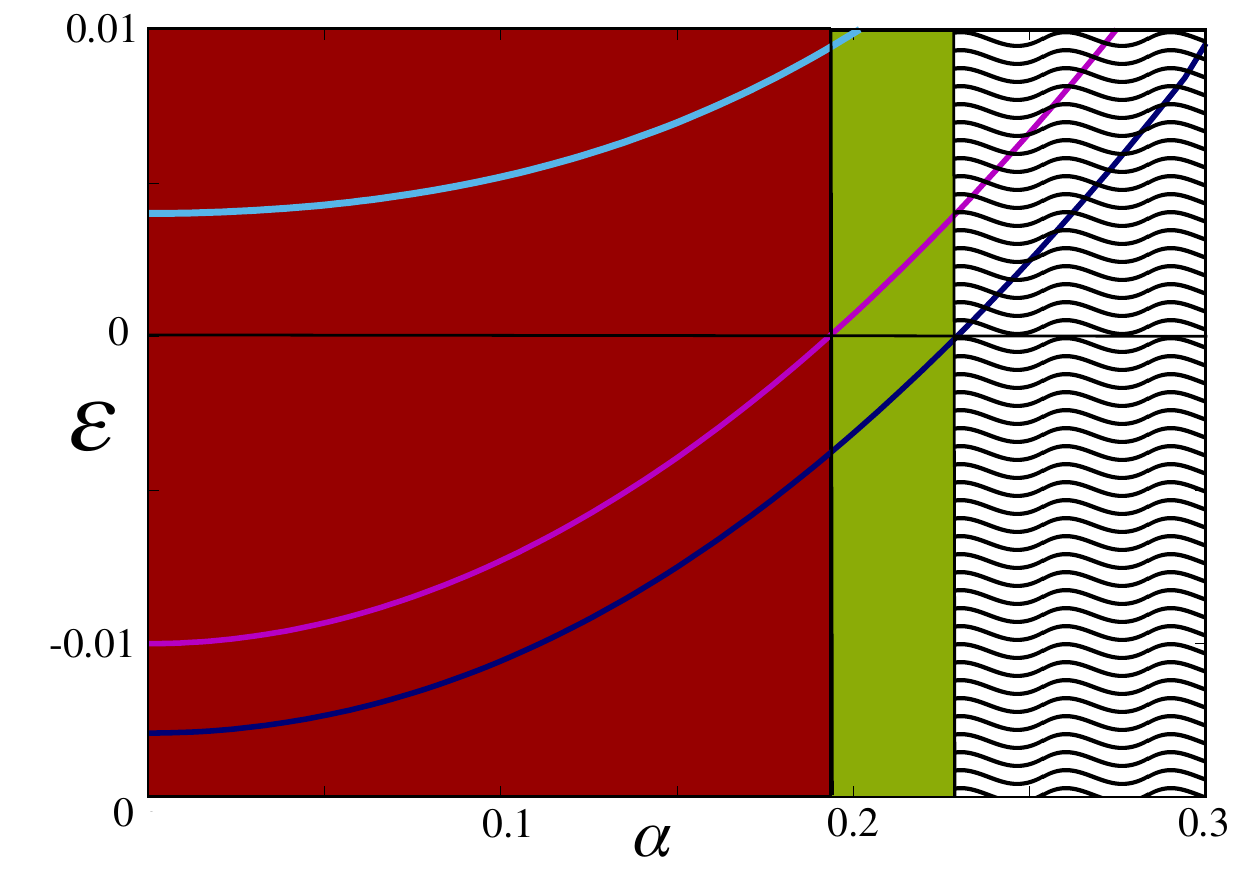}
                \caption{$LE_z=0.01t$, $\mu=0.02t$ }
                \label{Fig:big_lif1}
        \end{subfigure}~
        \begin{subfigure}[b]{0.33\textwidth}
                \includegraphics[width=6cm]{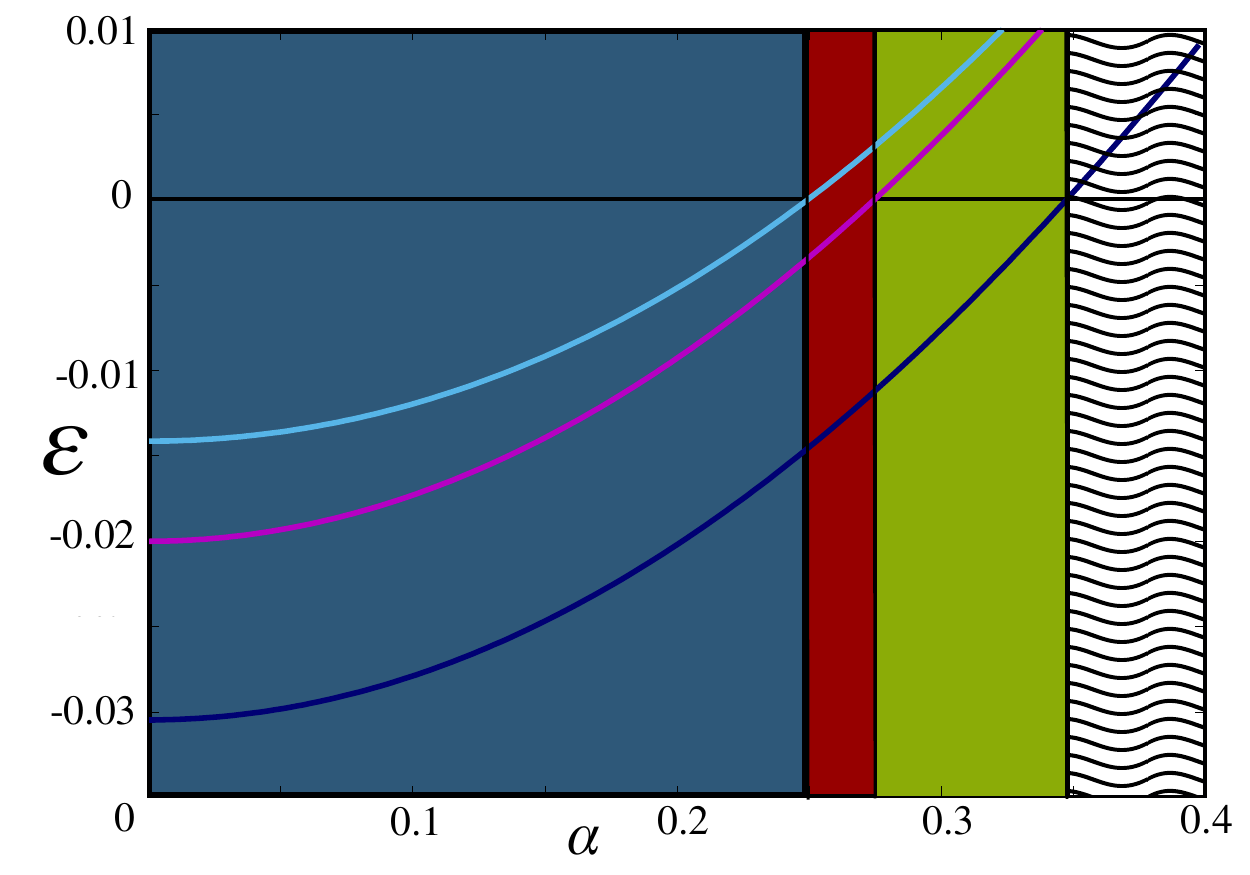}
                \caption{$LE_z=0.03t$, $\mu=0.05t$}
                \label{Fig:big_lif2}
        \end{subfigure}%
      \begin{subfigure}[b]{0.33\textwidth}
                \includegraphics[width=6cm]{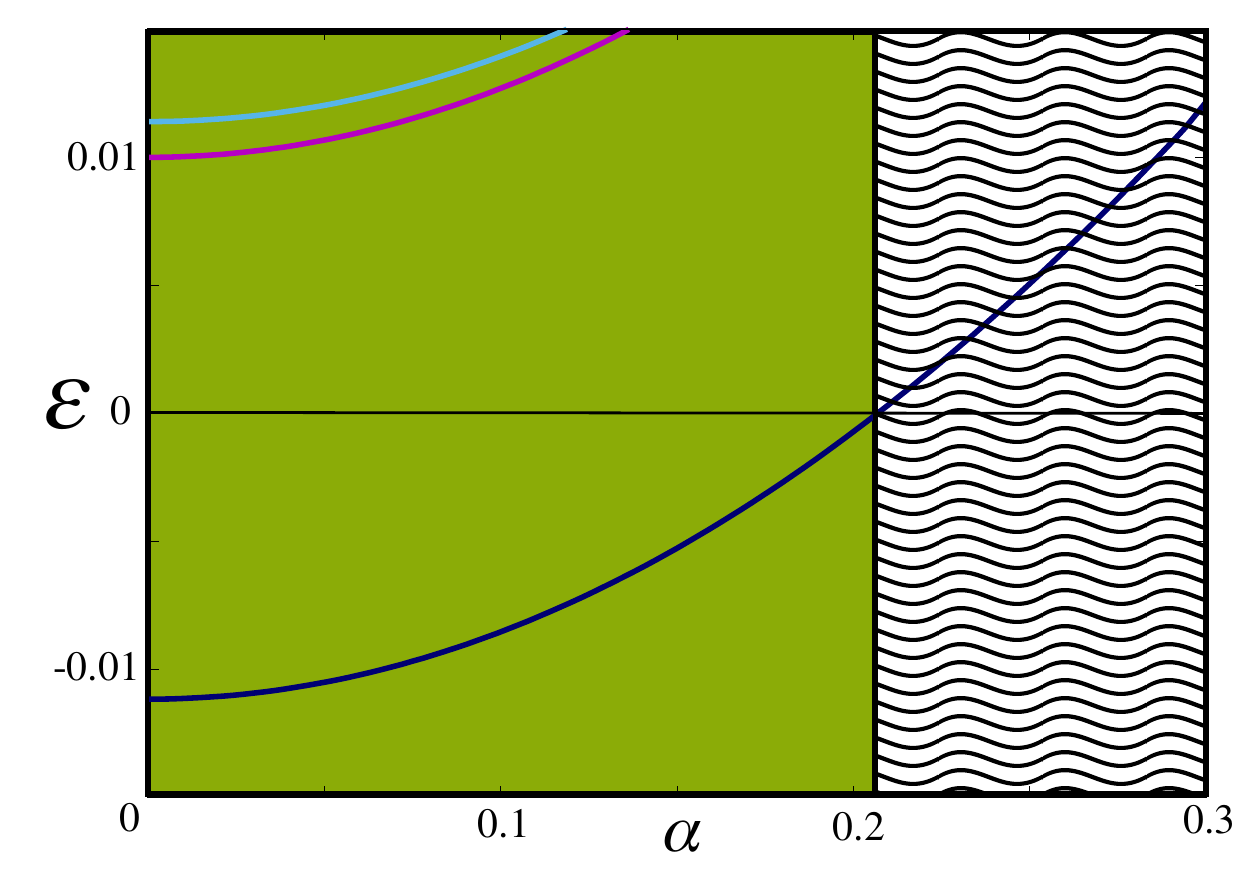}
                \caption{$LE_z=0.06t$, $\mu=0.05t$}
                \label{Fig:big_lif3}
        \end{subfigure}%
   \caption{ The Floquet Lifshitz phase diagram of bilayer graphene. The change in phases with increasing $\alpha$ is plotted for different initial values of $LE_z$ and $\mu$.
   The colors of the different phases are the same as that in Fig.\ref{Fig:bare_lifshitz}}
   \label{Fig:big_lif} 
  \end{figure*}

The effective Hamiltonian for bilayer materials in momentum space with renormalized interactions is given by
\begin{widetext} 
 \begin{align}
H_{\rm eff} =   \begin{pmatrix}
\Lambda^0_{q\sigma}+\Lambda^a_{q\sigma}+\Lambda^1_{q\sigma}+\mu^a_{q\sigma}  &\mathcal{J}_{q\sigma}+L_{q\sigma}+M_{q\sigma}&0& \tilde{T}_{3q\sigma}  \\
\mathcal{J}^\ast_{q\sigma}+L^\ast_{q\sigma}+M^\ast_{q\sigma}  & -\Lambda^0_{q\sigma}+\Lambda^b_{q\sigma}+\mu^b_{q\sigma}  &t_\perp&0  \\
 0&t_\perp & \Lambda^0_{q\sigma}+\Lambda^{\tilde{a}}_{q\sigma}+\mu^{\tilde{a}}_{q\sigma} &\mathcal{J}_{q\sigma}+L_{q\sigma}+M_{q\sigma} \\
 \tilde{T}^\ast_{3q\sigma} &0&\mathcal{J}^\ast_{q\sigma}+L^\ast_{q\sigma}+M^\ast_{q\sigma}  &-\Lambda^0_{q\sigma}-\Lambda^1_{q\sigma}+\Lambda^{\tilde{b}}_{q\sigma}+\mu^{\tilde{b}}_{q\sigma} \\
\end{pmatrix}
\label{Eqn:H_matrix2}
 \end{align}
  
\end{widetext}
where $T_3+T^\prime_3+T_4+T_5=\tilde{T}_{3q\sigma}$ and
\begin{eqnarray}
 \Lambda^{a/b}_{\sigma}&=&-\frac{t^2 \left( (L\pm l)E_z-\mu \right)}{\omega^2} \left(\sum_{n\ne 0} \frac{J^2_n(\alpha)}{n^2} \cos{\frac{2\pi n}{3}} \right) \nonumber\\
 \Lambda^{\tilde{a}/\tilde{b}}_{\sigma}&=&-\frac{t^2 \left( (-L\pm l)E_z-\mu \right)}{\omega^2} \left(\sum_{n\ne 0} \frac{J^2_n(\alpha)}{n^2} \cos{\frac{2\pi n}{3}} \right) \nonumber\\
 \mu^{a/b}_{\sigma}&=&\left(1-\frac{3t^2}{\omega^2} \sum_{n\ne 0} \frac{J^2_n(\alpha )}{n^2} \right)\left((L\pm l)E_z-\mu \right)\nonumber\\
 \mu^{\tilde{a}/\tilde{b}}_{\sigma}&=&\left(1-\frac{3t^2}{\omega^2} \sum_{n\ne 0} \frac{J^2_n(\alpha)}{n^2} \right)\left((-L\pm l)E_z-\mu \right)
\label{Eqn:lEzmu}
\end{eqnarray}
Using this effective Hamiltonian, we will now study the effect of driving at high frequency in both bilayer
graphene and bilayer silicene and obtain its effects in the next section.

\section{The phase diagrams}

In this section, we show how high frequency light can be used to obtain new phases and control changes in the Fermi surface topology, in
both bilayer graphene and bilayer silicene. For single layer materials, we already know that shining light leads to changes
in the Chern number and hence leads to several new phases.  Here, we wish to study changes in Chern number as well as changes in Fermi surface topology as a function of the amplitude of light. 

\subsection{Bilayer graphene}
 
The Hamiltonian in Eq.(\ref{Eqn:H_matrix2}) has terms involving spin-orbit coupling and a buckled lattice structure. 
These are set to zero for  bilayer graphene, $i.e.$, we set   $\tilde{T}_{3q\sigma}=T_3$, since there is neither buckling nor any spin dependence, and $l=0$, since there is no spin-orbit coupling,  along with $\mu=0$ in Eq.\ref{Eqn:lEzmu}.
This means that $\Lambda^a = \Lambda^b$ and $\Lambda^{\tilde{a}} = \Lambda^{\tilde{b}} $ with $\Lambda^a=-\Lambda^{\tilde{a}}$. Similarly
$\mu^a =\mu^b$ and $\mu^{\tilde{a}} = \mu^{\tilde{b}} $ with $\mu^a=-\mu^{\tilde{a}}$. The values of the inter-layer couplings are taken as $t_\perp=0.12t$ and $t_3=0.1t$.

 We  then expand the 
Hamiltonian around the $K$ and $K^\prime$ points and calculate the energy eigenvalues to study whether the trigonal warping is modified in the effective Hamiltonian. 
For simplicity, we set  $L E_z=0$ in the original undriven Hamiltonian. The position of the four gapless
Dirac cones then reduces to that given in Eq. \ref{Eqn:epsiloneta} with $LE_z=0$ which  simplifies the effective
B-W Hamiltonian in Eq. \ref{Eqn:H_matrix2}. The only  new remaining terms in the diagonal as compared to the static case,
$\Lambda^0, \Lambda^1$ are `effective spin-orbit couplings' and are, therefore, momentum independent in the low energy limit. Thus, even in the presence of driving, 
the functional form of the eigenvalues in Eq.\ref{Eqn:bigev1} does 
not change. However, the `effective spin-orbit coupling terms' introduced by driving leads to a mass gap,
given by $2|\Lambda^0+\Lambda^1|$
just like it does for genuinely spin-orbit coupled materials like silicene. Moreover, since this
gap is introduced by the driving, it is  dependent on the amplitude $\alpha$ of the light and  is
oscillating. These oscillations lead to gap closures and Chern number changes as shown in Fig.\ref{Fig:big_pd}.

We then compute  the Chern numbers for a range of values of $E_z$ and $\alpha$ for $\omega=10t$ to construct a phase diagram describing the various topological phases in this model
 shown in Fig.\ref{Fig:big_pd}. The phase diagram is symmetric under $LE_z \rightarrow -LE_z$in most parts of the phase diagram other than the regions near $\alpha=2.4$ and $5.5$, although the 
 terms $\Lambda^a$ and $\mu^a$ in the Hamiltonian 
change their sign under this symmetry.  However, since both
 these terms are very small compared to the other terms, this symmetry
 breaking is only visible when the all other terms are also small. 
 Other than that, the phase diagram shows the expected repetition of phases (with smaller
 areas)  as we increase the value of the amplitude $\alpha$ of light\cite{mohan2016}.

To study the Lifshitz transition using the B-W effective Hamiltonian, we consider three  different initial points in the Lifshitz phase diagram of bilayer graphene in Fig.\ref{Fig:bare_lifshitz} 
and study the effect of shining light on these phases. The inter-layer coupling values are taken as $t_\perp=0.5t$, $t_3=0.45t$ which are same as in Fig.\ref{Fig:bare_lifshitz}.
The  resultant phase diagrams as a function of the amplitude of light are depicted in Fig.\ref{Fig:big_lif}. The (a) magenta, (b) dark blue and (c) light blue curves are the heights (measured from $\varepsilon_F=0$ to the bottom of the cone)
of (a) the Dirac cone  at the $K$ point,
(b) the satellite Dirac cone and (c) the van-Hove singularity,  respectively.  The colours in this  phase
diagram are chosen to match the ones in the phase diagram without light  Fig.\ref{Fig:bare_lifshitz}.
In Fig.\ref{Fig:big_lif1}, in the red region, both the Dirac point  and the satellite Dirac points are below the Fermi level. The topology is that of four disconnected regions in the Fermi level. As the value of $\alpha$ increase from zero, close to $\alpha=0.2$,  the Dirac cone rises above the
Fermi level changing the topology to that of three disconnected regions. This is the green region in this phase diagram and in Fig.\ref{Fig:bare_lifshitz}. As $\alpha$ is further increased, the satellite cones arises above the Fermi 
level and we enter the insulating striped phase.  Thus, as a function of driving, we are able to tune a
Lifshitz (topology-changing) transition.
 
In Figs.~\ref{Fig:big_lif2} and \ref{Fig:big_lif3} we start from initial points in the 
blue and green phases in Fig.\ref{Fig:bare_lifshitz} and show the changes that occur under
driving  as shown in Fig.\ref{Fig:big_lif}. In Fig.\ref{Fig:big_lif2}, we start from the phase where 
even the van-Hove singularities  is below the Fermi level. By shining light,  we first move to a phase
with 4 disconnected regions (red phase) where the van-Hove singularities go above the Fermi level.
Further driving moves the Dirac point above the Fermi level and we reach the green phase
with three disconnected regions and finally, when even the satellite Dirac points go 
 above the Fermi level, we reach the insulating phase. In Fig.\ref{Fig:big_lif3}, we start from
 the green phase where only the three satellite Dirac points are below the Fermi level and 
 directly transition into the insulating phase.

\subsection{Bilayer spin-orbit coupled materials}
In this section, we study the Floquet phase diagram and the Lifshitz transition for spin-orbit coupled materials. 
 \begin{figure}[!ht]
   \includegraphics[width=9cm,height=6cm]{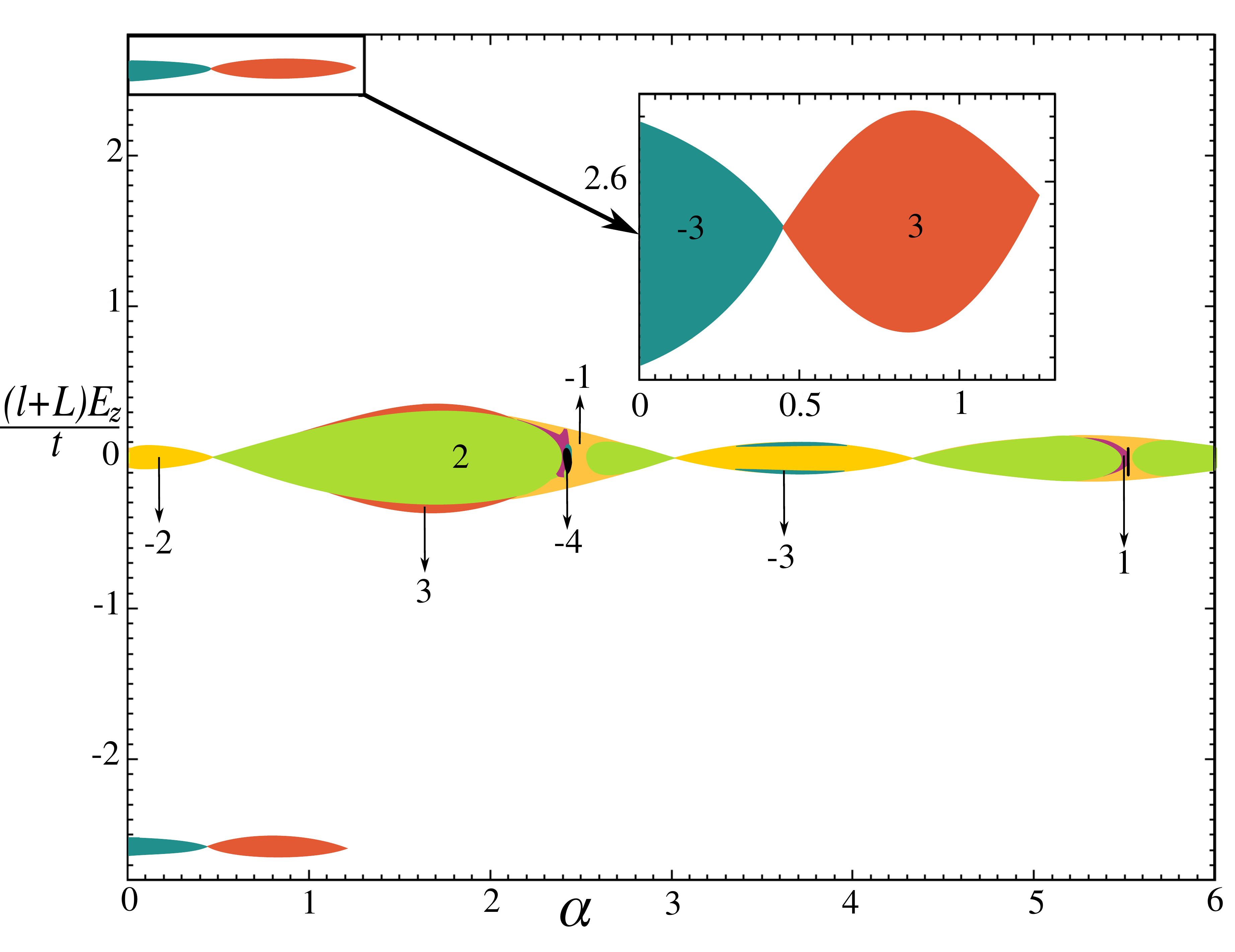}
   \caption{Floquet topological phase diagram of bilayer materials with spin-orbit coupling. The phase diagram is plotted for up spin. }
   \label{Fig:bisi_pd}
  \end{figure}
 The Floquet topological phase diagram for $\uparrow$ spin is calculated in Fig.\ref{Fig:bisi_pd} for $\lambda=0.05t$, $\omega=10t$, $t_\perp=0.12t$, $t_3=0.1t$ and $L=4l$ as a function of $E_z$ and $\alpha$. The phase diagram can be divided into three regions with 
  topological phases separated by trivial ones as we 
 increase the perpendicular electric field. The top and bottom regions corresponds to the $C_\uparrow=-3$ phase in the static phase diagram of silicene in Fig.\ref{Fig:bilayersi} and exists for both 
 positive and negative values of $E_z$. This phase continues to have a 
 non-zero Chern number $(3,-3)$ until
 $\alpha$ is increased to $\approx1.25$ before disappearing as shown in the inset. The middle region is similar to that of bilayer graphene in Fig.\ref{Fig:big_pd} and to that of the
topological phase diagram of irradiated single layer silicene studied using B-W theory\cite{mohan2016}. The $C_\uparrow=-2$ phase from Fig.\ref{Fig:bilayersi} extends for a small region in $\alpha$ while $C_\uparrow=-1$ region 
 is too narrow to be visible in a large phase diagram. 
  \begin{figure}[ht!]
    \centering
          \includegraphics[width=8cm, height=8cm]{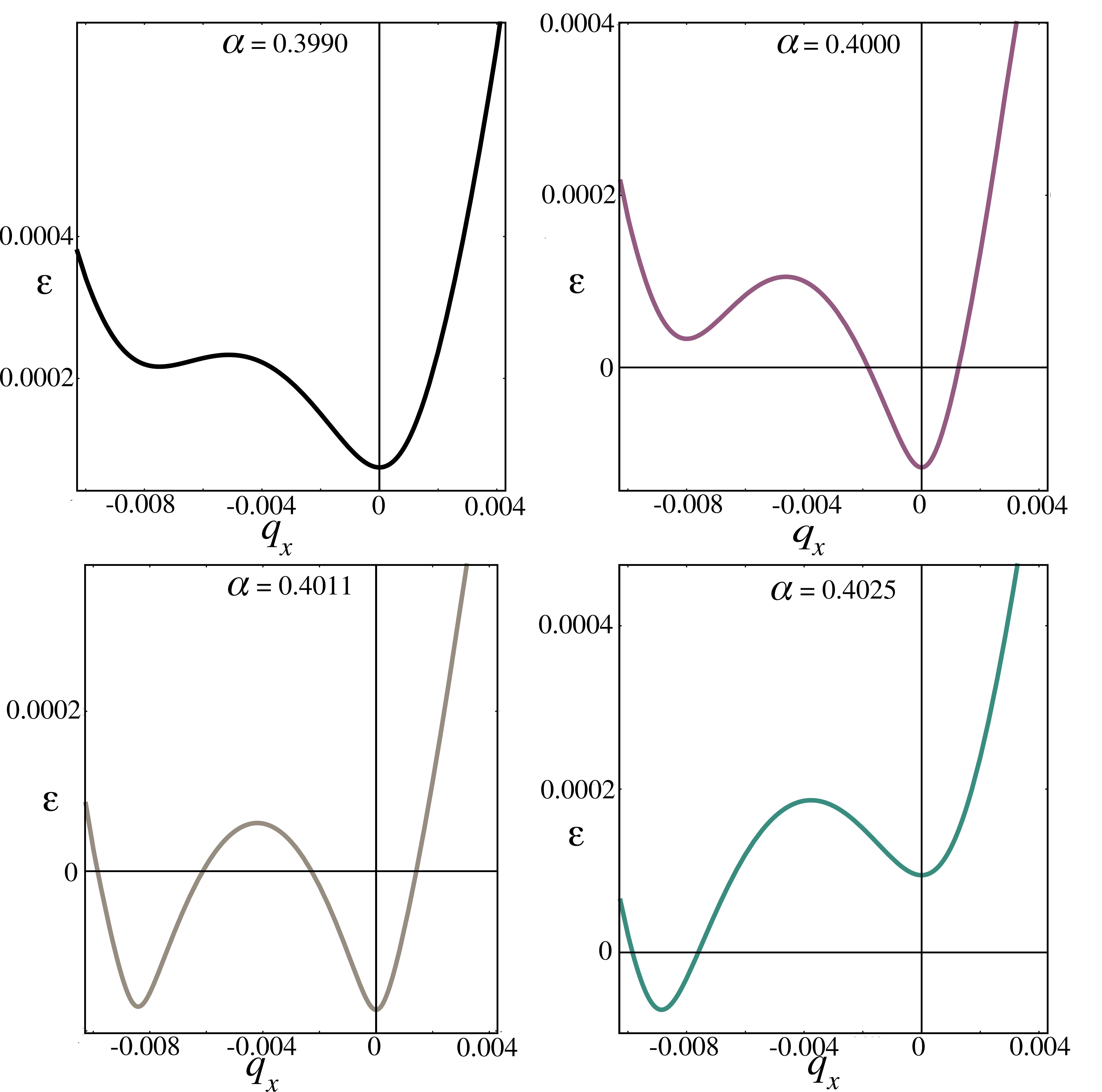}
       \caption{Floquet Lifshitz phase diagram for bilayer silicene with varying $\alpha$ for $(L+l)E_z= -0.01t$ and $\mu= 0.00025 t$. }
     \label{fig:bisi_lifshitz}
     \end{figure}
   
 In the rest of this section,  we try to understand the effect of an additional tuning parameter $\alpha$ on the Lifshitz phase transition of bilayer silicene.
 To compare it with the bare case described in Fig.\ref{figbisi}, we fix the values of $\mu$ and $E_z$ and tune the value of $\alpha$ as 
 shown in Fig.\ref{fig:bisi_lifshitz}. For $\mu/t= 0.00025$ and $(L+l)E_z/t = -0.01$, we have considered four values of $\alpha$ in Fig.\ref{fig:bisi_lifshitz}.
 In the four different panels, the color of the curve is same as the color of the  Lifshitz phase in Fig.\ref{figbisi} for that particular value of $\alpha$.
 The top left panel where $\alpha=0.3990$ corresponds to the striped phase in Fig.\ref{figbisi}. This is an insulating phase where the Fermi energy is in the gap.
 The top right panel with $\alpha=0.4000$ and the purple curve corresponds to the phase  with the same color in Fig.\ref{figbisi}. Here the electron pocket at the Dirac $K(K')$
 point is below the Fermi level and the topology is that of a disc corresponding to a  single region in the Fermi surface. The bottom left and the bottom right panels with $\alpha=0.4011$ 
 and $\alpha=0.4025$ correspond to the brown and green colored phases respectively in Fig.\ref{figbisi}.  The brown colored phase has the electron pockets 
 at  the Dirac point and at the three satellite points below the Fermi level. This implies that the 
 the Fermi surface is made up of four disconnected regions. The 
 green colored region consists of three disconnected regions since the electron pocket at the satellite points are below the Fermi level. Hence, as a function of the drive,
 we can tune the bilayer silicene through topology changing Lifshitz transitions - from an insulating
 phase in (a) to a single region Fermi surface in (b) to a four region Fermi surface in (c) and
 finally to a three region Fermi surface in (d). Note also that these transitions are extremely sensitive
 to the value of the amplitude of the drive and occur for very small changes in $\alpha$.   This is not
 too surprising because these changes are correlated with  changes in the Chern number at the same
 value of $\alpha$ in Fig.\ref{Fig:bisi_pd}, and in that figure, the region of the phase change from $-2$ to $-1$ to $0$ is so small that it cannot be shown in the figure, where it looks like the phase change is directly from $-2$ to $0$.

 \section{Discussion and conclusions}
 
 In this paper, we have studied the effect of light on bilayer graphene and silicene and seen how they affect
 the Fermi-surface topology changing  Lifshitz transitions.
 Physically, it is the magnetotransport properties, such as the Shubnikov-de Haas effect and thermodynamic properties
 such as the  de Haas-van Alphen effect  which are affected  by the changes in Fermi surface topology.  Essentially, when there are either creation of additional Fermi surface pockets or merging of Fermi surface pockets, the Landau level degeneracy changes.  For instance, when the Fermi level decreases  from the blue region   in Fig.\ref{Fig:bare_lifshitz}
 with a single Fermi surface to the green region, with three Fermi surface pockets, the period of the oscillations triple.
 So changes in the degeneracy which are caused either by the Lifshitz transitions  can be easily measured by the Shubnikov-de Haas  oscillations period.
 Moreover,  precisely at the multicritical point (black dot)  in Fig.\ref{Fig:bare_lifshitz} where the  critical points merge, the density of states diverges. This is also something which can be seen experimentally. 
 Finally,  in this paper, we have also  shown how these phases
 in bilayer graphene and in spin-orbit coupled  materials like bilayer silicene, can be controlled by shining light on these systems, paving the way for opto-electronic devices in these materials.

During the preparation of this paper for publication, we became aware of the work of Ref.\onlinecite{iorsh2017} who also study
the effect of light on Lifshitz transition in bilayer graphene.  Our results agree where there is overlap.

 \acknowledgements
 We would like to thank Abhishek Joshi, Arijit Kundu  and Ganpathy Murthy for useful discussions. PM would like to thank HRI for hospitality during the completion of this work.

\bibliographystyle{apsrev}

\end{document}